# Negative Poisson's ratio in graphene-based carbon foams

Jin Zhang[1*], Qilin Xiong[2,3*]

[1] Shenzhen Graduate School, Harbin Institute of Technology, Shenzhen 518055, China

[2] Department of Mechanics, Huazhong University of Science & Technology, 1037 Luoyu Road, Wuhan 430074, China

[3] Hubei Key Laboratory of Engineering Structural Analysis and Safety Assessment, 1037 Luoyu Road, Wuhan 430074, China

**Abstract:** Using molecular dynamics simulations, we find an in-plane negative Poisson's ratio intrinsically existing in the graphene-based three-dimensional (3D) carbon foams (CFs) when they are compressed uniaxially. Our study shows that the negative Poisson's ratio in the present CFs is attributed to their unique molecular structures and triggered by the buckling of the CF structures. This novel mechanism makes the negative Poisson's ratio of CFs strongly depend on their cell length, which offers us an efficient means to tune the negative Poisson's ratio in nanomaterials. Moreover, as the buckling modes of CFs are topographically different when they are compressed in different directions, their negative Poisson's ratio is found to be strongly anisotropic, which is in contrast to the isotropic positive Poisson's ratio observed in CFs prior to the buckling. The discovery of the intrinsic negative Poisson's ratio in 3D CFs will significantly expand the family of auxetic nanomaterials. Meanwhile, the novel mechanism of nano-auxetics proposed here may open a door to manufacture new auxetic materials at the nanoscale.

---

[*]Corresponding authors. J.Z. and Q.X. contributed equally to this work.

E-mail address: jinzhang@hit.edu.cn (J. Zhang); xiongql@hust.edu.cn (Q. Xiong).

Since the successful synthesis of low-dimensional carbon allotropes, such as zero-dimensional fullerenes, one-dimensional (1D) carbon nanotubes and two-dimensional (2D) graphene, the last several decades have witnessed an increasing interest in these novel carbon-based nanomaterials (CNMs). Studies show that these low-dimensional CNMs possess numerous unique physical and chemical properties superior to their conventional three-dimensional (3D) counterparts, e.g., graphite and diamond [1, 2]. Inspired by the observation of exceptional properties in these low-dimensional CNMs and the development of the synthesis method for nanomaterials, researchers recently are trying to use these low-dimensional CNMs as building blocks to construct novel 3D nanomaterials. In 1992 Karfunkel and Dressler [3] theoretically predicted a honeycomb-like 3D carbon foam (CF) structure, which is composed of the elements of graphene sheets. In these CF structures, each component graphene sheet holds the $sp^2$ carbon bond, while every three neighbouring graphene sheets are linked together by a line of $sp^3$-bonded carbon atoms [3-6]. The structural stability of these CF materials has been proven by numerous studies [3-6]. However, these CF structures did not receive their due attention until 2016 when Krainyukova and Zubarev [7] successfully fabricated these CFs in their experiments. The experimental observation of 3D CFs has triggered growing studies on the physical properties of CFs. Studies reveal that these honeycomb CF structures possess numerous extraordinary properties, e.g., high thermal conductivity [6, 8-10], large mechanical strength [4-6] and high level of physical absorption [7, 9, 10]. These unique properties render CFs attractive in various engineering applications, such as gas storage [7, 9, 11, 12], energy absorption [13], etc.

In addition to the honeycomb CF structures described above, researchers also proposed a new type of CF [4], i.e., the triangular CF shown in Fig. 1a. In this triangular CF, at each joint six adjacent graphene sheets are connected together by a hexagon carbon ring. Different to the honeycomb CF which contains mixed $sp^2$-$sp^3$ carbon bonds, the present triangular CF uniquely contains $sp^2$ carbon bonds only, which is thus considered to be a unique 3D CNM [4]. The structural stability of these triangular CF structures has been confirmed by Wu et al. [4] in their density-functional theory (DFT) calculations, which strongly proves the possibility of synthesizing these triangular CF structures in experiments. It is highly expected that the novel fabrication technique developed recently for the honeycomb CFs [6, 14] can be extended to fabricate the triangular CFs in the near future. Although the physical properties of honeycomb CFs have been well studied recently [4-10, 12, 13], these results may be not applicable for their triangular counterparts because honeycomb and triangular CFs possess vastly different molecular structures. Indeed, Wu et al. [4] reported that the Young's modulus of triangular CFs is greatly larger than that of their honeycomb counterparts. This pioneering study to some extent denotes that the triangular CFs may have some unique mechanical properties when comparing to their honeycomb counterparts.

Motivated by these ideas, this paper aims to achieve a comprehensive understanding of the fundamental mechanical properties of triangular CFs. In doing this, molecular dynamics (MD) simulations are performed to measure the mechanical properties including Young's modulus and Poisson's ratio of CFs with different cell lengths. A large Young's modulus is detected in the present triangular CFs, which is also found to depend on their cell length. More importantly, we find a negative Poisson's ratio in CFs leading to contraction in transverse directions when they are under uniaxial compressive loads. Theoretical analyses are also conducted to reveal the physical mechanisms behind these unique behaviours.

In the present study, classical MD simulations are adopted to investigate the mechanical behaviours of triangular CFs. Specifically, in the simulations the interactions between carbon atoms in CFs are described by the intermolecular reactive empirical bond order (AIREBO) potential [15]. This AIREBO potential has been widely utilized in investigating the mechanical behaviours of various CNMs including carbon nanotubes [16, 17], graphene sheets [18, 19] and honeycomb CFs [5, 12, 13]. MD results based on this AIREBO potential are proven to be in good agreement with experimental data [20] and also results of density functional theory (DFT) simulations [21]. In each simulation, the molecular structure of a triangular CF structure is initially built using the lattice constant calculated based on first-principles [4] and subsequently relaxed to a minimum energy state using the conjugate gradient algorithm. The achieved energy-minimized structures of CFs are thus ready for MD simulations of the mechanical behaviours. In doing this, after the periodic boundaries are applied in all directions (i.e., $x$, $y$ and $z$ directions), CFs are relaxed for a certain period (100000 time steps) to minimize the internal energy and reach an equilibrium state at the temperature of 1 K and pressure of 0 Pa. In this process, the barostat (*NPT*) ensemble and Nosé-Hoover thermostat algorithm [22] are utilized to maintain the constant temperature and pressure. The system is then further regarded as a microcanonical ensemble (*NVE*) for 50000 time steps to check the equilibrium state. Here, we set the system temperature as 1 K because the influence of thermal fluctuation is extremely small at such a low temperature. In addition, the velocity Verlet algorithm with a time step of 1 fs is utilized to integrate the Hamiltonian equations of motion determined by Newton's second law. After the full relaxation, the CF structures are quasi-statically loaded under a uniaxial compressive load along $x$ or $y$ direction as shown in Fig. 1a. In doing this, we apply the compressive mechanical strain along one direction ($x$ or $y$ direction) of CFs, while the CF structure is allowed to fully relax in the transverse direction. In this process, periodic boundary conditions are also applied in the transverse direction, while both ends in the axial direction are translated rigidly to apply the tensile deformation to the CFs. Moreover, to simulate the quasi-static load applied on the CF structures in simulations we choose a relatively low strain rate of 0.01 ns$^{-1}$. In the present study, all MD simulations are conducted using a large-scale atomic/molecular massively parallel simulator (LAMMPS) [23].

The MD techniques detailed above are employed to study the mechanical behaviours of triangular CFs under uniaxial compressive loads. To study the influence of the length of component graphene elements on the mechanical behaviours of entire CF structures, in the present study we consider different CFs whose cell length $d$ (or the length of the component graphene element) shown in Fig. 1a varies from 1.3 nm to 2.8 nm. Using the CASTEP package, DFT calculations are performed to confirm the structural stability of these triangular CFs. The generalized gradient approximation of the Perdew-Burke-Ernzerhof functional [24] is used for the DFT computation conducted here. No structural reconstruction is observed in the CFs upon structural relaxation, denoting the structural stability of the present triangular CFs. Moreover, our DFT calculations also show that triangular CFs are metallic with zero band gaps (see Fig. 1b), irrespective of their cell length.

From Fig. 1a we see that CFs show different configurations in $x$ and $y$ directions, suggesting that the CF could be an anisotropic structure with different mechanical properties in these two directions. Motivated by this idea, we will study the influence of the configuration on the mechanical responses of CFs by uniaxially compressing CFs in $x$ and $y$ directions, respectively. In Figs. 2a and 2b we show the stress-strain relation for CFs when they are compressed along $x$ and $y$ directions, respectively (see Figs. 2c and 2d). Here, the strain $\varepsilon$ for CFs is calculated by $\varepsilon = (L - L_0)/L_0$, where $L$ and $L_0$ are the lengths of CFs after and before the deformation, respectively. The stress $\sigma$ is calculated by $\sigma = (\partial U/\partial \varepsilon)/V$, where $U$ is the strain energy stored in the deformed CFs and $V$ is the volume of CF structures. From Figs. 2a and 2b we see that at the small deformation the stress in CFs linearly increases as the strain increases, corresponding to a linear elastic deformation in this process. Thus, according to the well-known Hooke's law, the corresponding Young's modulus $E$ of CFs can be calculated as the slope of the stress-strain curve in the linear elastic deformation stage. In Fig. 3 (symbols) we show the Young's modulus of CFs with different cell lengths $d$ calculated from MD simulations. The results show that the difference of Young's modulus in $x$ and $y$ directions is extremely trivial, which indicates that the present CFs are elastically isotropic. From Fig. 3 we also find that the Young's modulus of CFs significantly drops as the cell length increases. For example, when the cell length is 1.3 nm the Young's modulus of CFs is around 275 GPa, which is extremely close to the value of 288 GPa obtained in previous DFT calculations [4] for CFs with the same geometric size. When the cell length grows to 2.8 nm the Young's modulus is found to decrease by 51% and become 135 GPa. The isotropic and cell length-dependent Young's modulus of CFs can be well understood by the continuum mechanics analysis. According to the continuum model of CFs we detailed in the Supplementary Information the Young's modulus $E_x$ and $E_y$ in $x$ and $y$ directions of CFs with different cell lengths $d$ can be approximately expressed as

$$E_x = E_y = \frac{2Y}{\sqrt{3}} \frac{1}{d}, \tag{1}$$

where $Y$ is the tensile rigidity of graphene. In Fig. 3 we find our MD results can be well fitted by Eq. 1. Both MD calculations and theoretical predictions prove the significant dependence of the Young's modulus of CFs on their cell length, which indicates that the elastic properties of CFs can be efficiently tailored by adjusting their cell length. Moreover, as a cellular nanomaterial with relatively low mass density $\rho$ [4], it is expected that CFs could have high specific strength, i.e., $E/\rho$. Indeed, in the inset of Fig. 3 we see that the specific strength of the present triangular CFs is around $149\,\text{GPa}\cdot\text{cm}^3/\text{g}$, which is about three times larger than the value of their honeycomb counterparts [6]. Moreover, in Fig. 3 the specific strength of CFs is found to be almost independent on the cell length. The cell length-independent specific strength of CFs can be attributed to the fact that, similar to their Young's modulus, the mass density of CFs is also proportional to the inverse of the cell length, i.e., $\rho \propto 1/d$, since in the continuum model we detailed in the Supplementary Information the mass density is approximately $\rho \approx 2\sqrt{3}P/d$, where $P$ is the areal mass density of graphene.

From the stress-strain relation shown in Fig. 2 we can also see that when the strain is larger than a certain critical value the stress will almost keep unchanged with the increase of the strain, resulting in a stress plateau stage in the stress-strain curve shown in Fig. 2. The appearance of the stress plateau stage is attributed to the buckling in the CFs, which will topographically change the configuration of CFs as shown in Fig. 4 and thus will greatly reduce the ability of CFs to withstand in-plane compressive loading. Here, the strain at the onset of CF buckling is defined as the critical buckling strain $\varepsilon_{cr}$, and the associated stress is referred to as the critical buckling stress $\sigma_{cr}$. On the other hand, the stable stress plateau in the present CFs renders them appealing in the energy absorption applications [13, 25]. Specifically, CFs with smaller cell length have larger energy absorption ability, since the energy absorption per unit volume of CFs is approximately $W \approx \sigma_{cr}\varepsilon$, where $\sigma_{cr}$ shown in Fig. 2 is found to increase as the cell length decreases.

Next we will shift our focus on the Poisson's ratio of CFs. In Fig. 5 we show the transverse strain and the Poisson's ratio against the axial strain of CFs with different cell lengths when they are subjected to the compressive loading along *x* and *y* directions, respectively. Here the Poisson's ratio is calculated as [26] $v_{ij} = -\partial \varepsilon_j / \partial \varepsilon_i$ ($i, j = x$ or $y$), where $\varepsilon_i$ and $\varepsilon_j$ are the axial strain and the transverse strain, respectively. From Figs. 5a and 5b we see that when the CFs is under a relatively small deformation the transverse strain of CFs has a sign opposite to its axial strain, irrespective of the loading direction. Moreover, under this condition the magnitude of the transverse strain is found to increase with increasing axial strain, resulting in a positive Poisson's ratio in CFs as shown in Figs. 5c and 5d. The Poisson's ratio of CFs under small deformation is found to range between 0.32 and 0.37 both in *x* and *y* directions, indicating that at small deformation CFs possess an isotropic Poisson's ratio. The isotropic Poisson's ratio ranging from 0.32 to 0.37 calculated from the present MD simulations is found to be in good accordance with $v_{xy} = v_{yx} = 1/3$ predicted from our continuum model detailed in the Supplementary Information. Further raising the axial strain up to a certain critical value, we find that the transverse strain turns to decrease with growing axial strain and eventually becomes negative, corresponding to the contraction in the transverse direction when the CFs are compressed axially. This result indicates that after the critical axial strain the Poisson's ratio of CFs shifts from a positive value to a negative value, which is confirmed by the MD results shown in Figs. 5c and 5d. Comparing Fig. 5 to Fig. 2 we find that the critical axial strain corresponding to the sign change of the Poisson's ratio is same as the value of the critical buckling strain, which signifies that the negative Poisson's ratio observed in CFs is triggered by the buckling of CFs. Moreover, it is noted that only CFs with the present triangular configuration are found to possess the negative Poisson's ratio. In our previous studies on the honeycomb CFs no negative Poisson's ratio was observed, even after their buckling. The negative Poisson's ratio observed in the present triangular CFs render it appealing for the use in absorbing sound and vibration, enhancing indentation resistance and so on [27, 28].

To better understand the mechanism of the negative Poisson's ratio in the present triangular CFs induced by the buckling, we will analyse the configuration of CFs after buckling. In Figs. 4c and 4d we show the unit cells of the buckled CFs when they are compressed along $x$ and $y$ directions, respectively. Moreover, to facilitate the following discussion we choose a representative component graphene sheet from the unit cells and note it as sheet $ab$ in Figs. 4c and 4d. It can be seen from Figs. 4c and 4d that the buckling of CFs will induce the rotation of the central hexagon carbon ring. In response to this rotation, the entire structure of sheet $ab$ which is connected to the hexagon carbon ring through atom $a$ will move inward. Moreover, in addition to the movement of the entire structure of sheet $ab$, the rotation of the central hexagon carbon ring will also induce the bending of sheet $ab$, which makes it transfer from a flat structure to a curved structure. As a result, the effective distance between the two ends of sheet $ab$, i.e., the distance between atoms $a$ and $b$ will become smaller due to the bending deformation. The above two deformations (inward movement and bending) of the component graphene sheets induced by the buckling of CFs will both contribute to the contraction of CFs in the transverse contraction and thus the negative Poisson's ratio in the present CFs after buckling.

The mechanism of the negative Poisson's ratio in the present triangular CFs depicted above can be utilized to explain some unique phenomena observed in the negative Poisson's ratio of the triangular CFs. For example, as the negative Poisson's ratio behaviour in the present triangular CFs is triggered by their buckling inability, the negative Poisson's ratio will only exist in the post-buckling process of CFs under compression, which usually corresponds to nonlinear elastic deformations in CFs. Due to the nonlinear deformations in the post-buckling process, in Figs. 5c and 5d the negative Poisson's ratio of CFs is found to depend on the axial compressive strain. As for CFs which are compressed along the *x* direction and have a cell length of 1.3 nm, in Fig. 5c their Poisson's ratio $v_{xy}$ is found to increase from -0.15 to -0.35 when the axial compressive strain $\varepsilon_x$ increases from 0.023 (critical buckling strain) to 0.2. As for CFs having the same cell length but compressed along the *y* direction, we find in Fig. 5d that as the axial compressive strain $\varepsilon_y$ increases from 0.022 (critical buckling strain) to 0.2 the Poisson's ratio $v_{yx}$ increases from -0.28 and reaches a maximum value of -0.46 at $\varepsilon_y = 0.06$. After that, the Poisson's ratio $v_{yx}$ decreases gradually with rising axial compressive strain $\varepsilon_y$ and reaches -0.35 when $\varepsilon_y$ grows to 0.2. The different relationships between the negative Poisson's ratio and the axial strain in *x* and *y* directions can be attributed to the different buckling modes of CFs when they are compressed in these two different directions as shown in Fig. 4. The different buckling modes of CFs compressed respectively in *x* and *y* directions also make the negative Poisson's ratio of CFs quantitatively different in these two directions. For instance, as for CFs whose cell length is 1.8 nm, when the axial compressive strain $\varepsilon_x$ or $\varepsilon_y$ is 0.2, $v_{xy}$ of the CFs is -0.16 which is 43% smaller than the value -0.28 of its $v_{xy}$ counterpart. This result shows that the negative Poisson's ratio in CFs induced by the buckling is strongly anisotropic, which is in contrast the isotropic Poisson's ratio detected in CFs prior to the buckling instability. Moreover, from Figs. 5c and 5d we also see that the negative Poisson's ratios in *x* and *y* directions both decreases as their cell length *d* increases. For example, when *d* increases from 1.3 nm to 2.8 nm $v_{xy}$ and $v_{yx}$ of CFs under an axial strain of 0.2 is found decrease from -0.35 to -0.05 and from -0.33 to -0.23, respectively. This result suggests that the negative Poisson's ratio in CFs can be efficiently adjusted by changing their cell length. To explain this cell length-dependent negative

Poisson's ratio observed in the present CFs we will refer to the buckling modes shown in Figs. 4c and 4d again. After CF buckling the inward movement of sheet *ab* induced by the rotation of the central hexagon carbon ring is determined by the side length *l* of the central hexagon carbon ring (see Fig. 1). which equals to the C-C bond length and is identical for all CFs with various cell lengths. Accordingly, the inward movement of sheet *ab* induced by the rotation of the central hexagon carbon ring is also identical for all CFs with various cell lengths. On the other hand, as for CFs with different cell lengths their geometric size will increase as the cell length increases, resulting in the decrease of the transverse strain with increasing the cell length of CFs. As a result, the Poisson's ratio of CFs after their buckling is expected to decrease as the cell length increases, which is in consistent with our MD results. Based on this theoretical analysis, we can further estimate the negative Poisson's ratio of CFs with different cell lengths. Taking CFs which are compressed along the *y* direction for example, their negative Poisson's ratio $v_{yx}$ under a relatively large axial strain of 1 can be estimated by the geometrical predictions detailed in the Supplementary Information. $v_{yx}$ obtained from the geometrical predictions can be expressed as

$$v_{yx} = -0.2 \cdot \left( \frac{1 + 5l/d}{1 + 2l/d} \right). \tag{2}$$

From Eq. 2 we see that, as we predicted above, the negative Poisson's ratio of CFs indeed decreases as the cell length increases. Specifically, when the cell length is relatively large, $v_{yx}$ predicted from Eq. 2 ultimately becomes -0.2, which is in good agreement with MD results shown in Fig. 5d.

From the above discussion, it is clear that the buckling of CFs plays a key role in initiating the negative Poisson's ratio in CFs. Thus, it is of great importance to understand the buckling behaviours of CFs. In Fig. 6a we plot the critical buckling stress of CFs with different cell lengths. We can see from Fig. 6a that the critical buckling stress decreases as the cell length increases, irrespective of the direction of the load applied on CFs. For example, when the CFs are compressed along the *x* direction, their critical buckling stress is found to drop from 6.0 GPa to 0.4 GPa when the cell length increases from 1.3 nm to 2.8 nm. In the same process, the critical buckling stress declines from 6.3 GPa to 0.7 GPa when CFs are compressed along the *y* direction. In fact, Fig. 4 shows that the buckling of CFs is triggered by the Euler buckling of their component graphene elements. According to the well-known Euler buckling theory [29] the critical buckling stress $T_{cr}$ of the component graphene elements can be expressed as

$$T_{cr} = \frac{n^2 \pi^2 D}{d^2}. \tag{3}$$

Here, *D* is the bending stiffness of graphene. *n* is the end constraint factor and ranges from 0.5 to 2 [25, 29], depending on the degree of constraint to rotation at two ends of the component graphene sheets. After applying the stress analysis to the unit cell of the continuum model of CFs shown in Fig. 6b we can approximately obtain the effective stresses $T_1$ in the inclined cell wall and $T_2$ in the horizontal cell wall of CFs when they are compressed in different directions (see the Supplementary Information for details). When CFs are subjected to a compressive stress $\sigma_x$ in the *x* direction, $T_1 = 0$ and $T_2 \approx \sqrt{3}\sigma_x d/2$, which indicates that in this case the buckling of CFs is induced by the buckling of their horizontal component graphene sheets. When CFs are subjected to a compressive stress $\sigma_y$ in the *y* direction, $T_1 \approx \sqrt{3}\sigma_y d/3$ and $T_2 \approx -\sqrt{3}\sigma_y d/6$, which indicates that in this case the buckling of CFs is induced by the buckling of their inclined component graphene sheets. It is noted that the stress distribution predicted from the present continuum model is in good accordance with the MD calculations shown in Figs. 2c and 2d.

Replacing $T_{cr}$ in Eq. 3 by $T_2$ when CFs are compressed along the $x$ direction and by $T_1$ when they are compressed along the $y$ direction, we can theoretically predict the critical buckling stresses $\sigma_{cr}$ of CFs when they are compressed in different directions, which are approximately $\sigma_{cr} \propto d^{-3}$ regardless of the loading direction. We find in Fig. 6a that our MD results can be well fitted by this theoretical prediction. Moreover, when $\sigma_x$ and $\sigma_y$ applied on CFs have the same magnitude, from the above analysis we see that $\sigma_x$ will induce a larger compressive stress in the component graphene sheets of CFs. Thus, it is reasonable to expect that the critical buckling stress of CFs compressed in the $x$ direction is smaller than that of CFs compressed in the $y$ direction. Indeed, this prediction is in good consistent with our MD results shown in Fig. 6a.

Finally, it is worth mentioning that since the first report of the negative Poisson's ratio behaviour in nanomaterials in 2014 for both black phosphorus [30] and metal nanoplates [26], numerous intrinsic and extrinsic mechanisms have been found or proposed to generate the negative Poisson's ratio (or so-called auxeticity) in nanomaterials [31]. Most reports of the auxeticity are in the 2D nanomaterials, e.g., black phosphorus [30], graphene [32-39], borophene [40, 41], 2D transition metal dichalcogenides [42] and 2D silica [43, 44]. However, very few reports of 3D nanomaterials have emerged. In the present study we report that recently proposed graphene-based 3D CFs intrinsically possess a large and tunable in-plane negative Poisson's ratio due to their unique molecular structures. The negative Poisson's ratio in the present 3D CFs is achieved through their mechanical instabilities (i.e., buckling). It is expected that the new mechanism of nano-auxetics detected in the present CFs can be extended to generate similar negative Poisson's ratio behaviours in other 3D network nanostructures made from the elements of 2D graphene sheets (or nanofilms) or 1D nanotubes (or nanowires). Moreover, recently similar buckling mechanism is also treated as a new means to achieve the negative Poisson's ratio in macroscopic materials [45-49], in which the negative Poisson's ratio is found to strongly depend on the structural configuration of materials. Thus, it is also highly expected that the macroscopic materials with a triangular configuration similar to the present CFs will have similar negative Poisson's ratio behaviours. This prediction is confirmed by our buckling analysis of a macroscopic foam structurally analogous to the present CFs. When the macroscopic foams are compressed along $x$ and $y$ directions, their buckling modes are shown in Figs. 7a and 7b, respectively, which are found to resemble to the structures of the buckled CFs shown in Fig. 4. Here, the buckling analysis is performed by using finite element (FE) calculations, which are carried out using the commercial code ANSYS. In this process, BEAM3 element is selected to describe the elastic wall of the foam structure. The FE model and the buckling analysis method of the foam structure are detailed in the Supplementary Information.

In summary, in this work MD simulations are carried out to study the mechanical properties of graphene-based CF structures under uniaxial compressive loading. A large Young's modulus is detected in the present CFs, which is also found to strongly depend on the cell length. More importantly, we find a negative Poisson's ratio in CFs leading to the contraction in transverse directions when they are under uniaxial compressive loads. This negative Poisson's ratio in CFs is attributed to their unique molecular structures and triggered by the buckling of the CF structures. Our results show that this new mechanism provides us an efficient means to tune the negative Poisson's ratio by adjusting the cell length of CFs. Moreover, as the buckling modes of CFs are topographically different when they are uniaxially compressed in different directions, their negative Poisson's ratio is found to be strongly anisotropic, which is in contrast to the isotropic Poisson's ratio observed in CFs prior to the buckling. The discovery of the intrinsic negative Poisson's ratio in the present 3D CFs will not only greatly expand the family of auxetic nanomaterials but also promote some new potential applications of CFs in novel nanodevices. Also, the new mechanism of mechanical instability proposed in the present study may open a door to manufacture new auxetic materials at the nanoscale.

**Acknowledgements**

This work was supported by the National Natural Science Foundation of China (Nos. 11602074 and 11502085). J.Z. acknowledges the financial support from Harbin Institute of Technology (Shenzhen Graduate School) through the Scientific Research Starting Project for New Faculty. Q.X. acknowledges the financial support from the Natural Science Foundation of Hubei Province in China (No. 2016CFB542) and the Fundamental Research Funds for the Central Universities (No. 2016YXMS097).

**Figures**

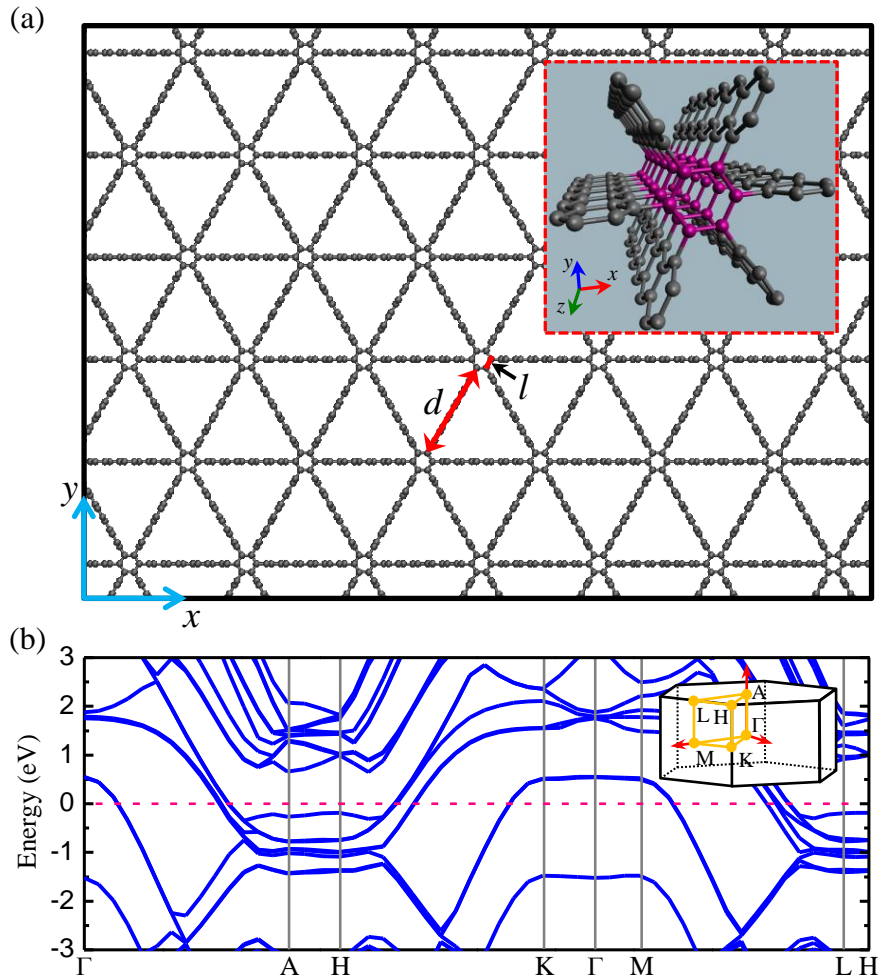

Figure 1. Structures of CFs. (a) Atomic representation of the CF structure. The CF is constructed by elements of zigzag-edged graphene sheets with a length of *d*. Each six adjacent graphene sheets are linked together by a line of sp$^2$-bonded carbon atoms, forming a hexagon carbon ring at the joint as shown in the inset. Here the side length *l* of the hexagon carbon ring equals to length of the carbon-carbon bond and is identical for all CFs with various cell lengths. (b) A representative band structure of CFs. The inset shows the first Brillouin zone.

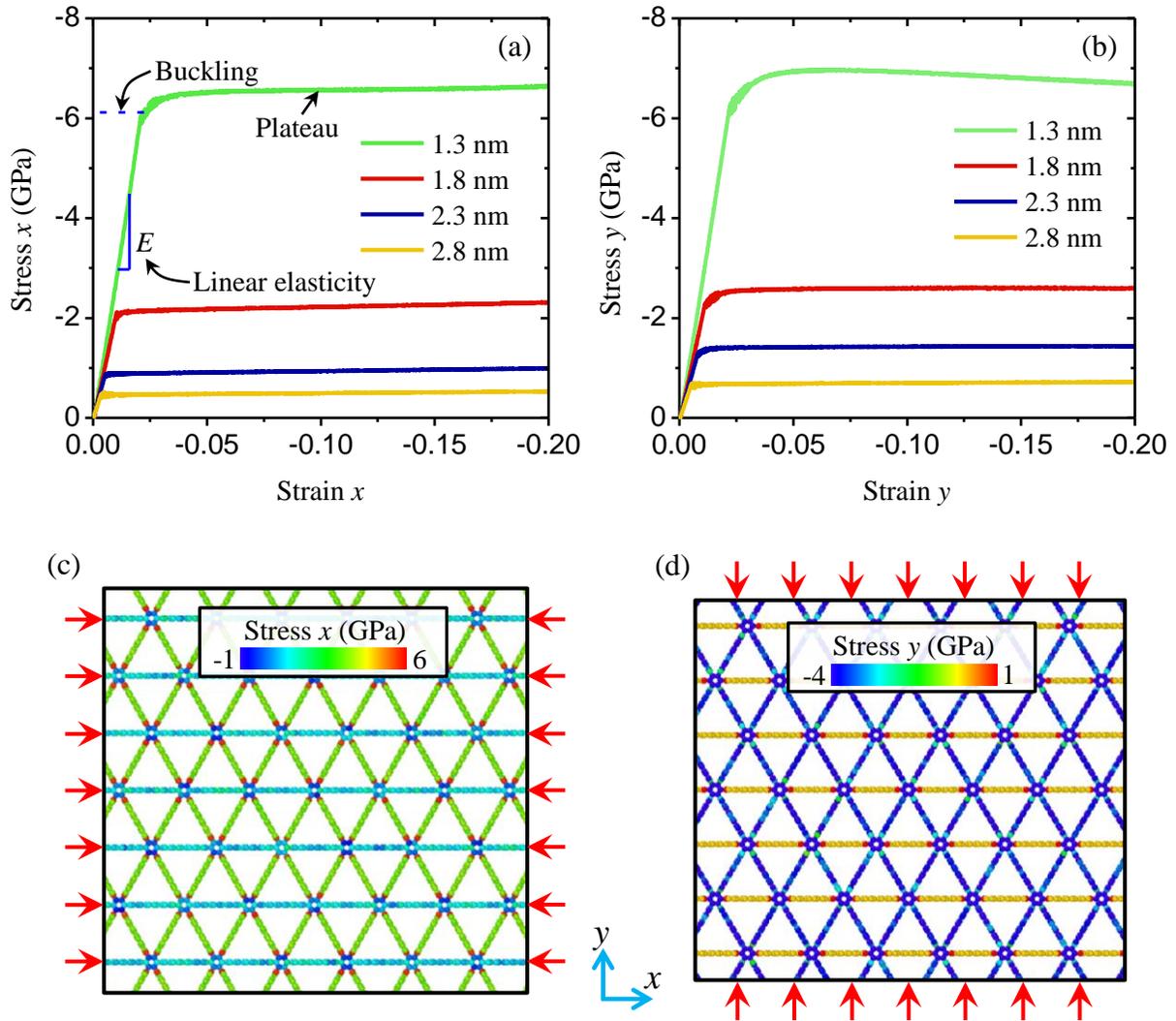

Figure 2. Mechanical responses of CFs subjected to the uniaxial compression. Top: stress-strain response of CFs subjected to a uniaxial compression along (a) *x* and (b) *y* directions. CFs experience linearly elastic deformations initially. When stress or strain grows beyond a critical value, CFs collapse by elastic buckling, resulting in a stable stress plateau stage in the stress-strain curve. Bottom: atomic stress distribution of CFs uniaxially compressed along (c) *x* and (d) *y* directions.

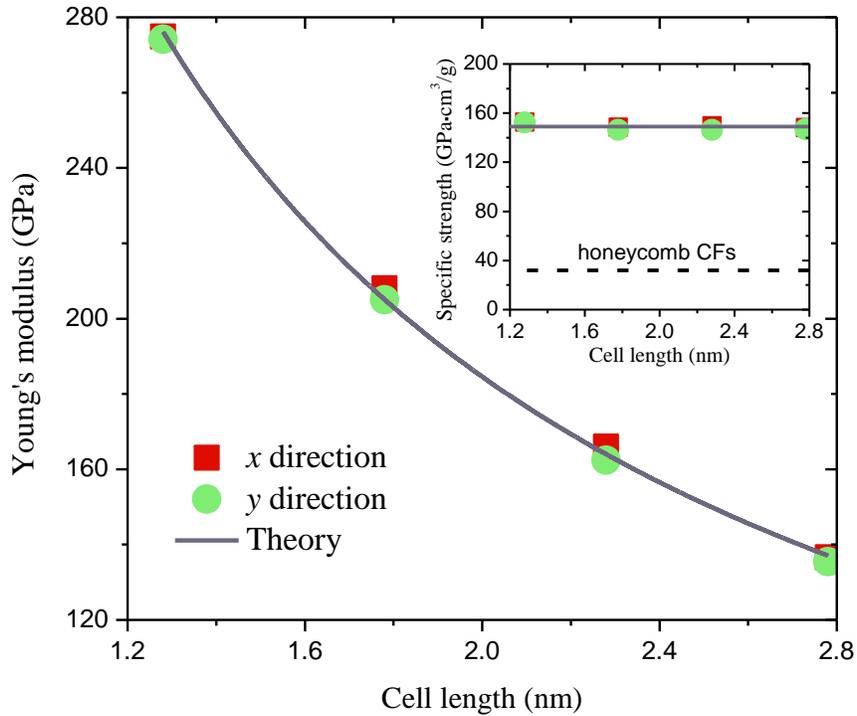

Figure 3. Young's modulus $E$ of CFs with different cell lengths $d$. Young's modulus in the $x$ direction is identical to that in the $y$ direction, irrespective of the cell length. Young's modulus of CFs calculated from MD simulations can be well fitted by $E \propto d^{-1}$ predicted from their equivalent continuum model. The inset shows the specific strength of CFs, which is independent with the cell length.

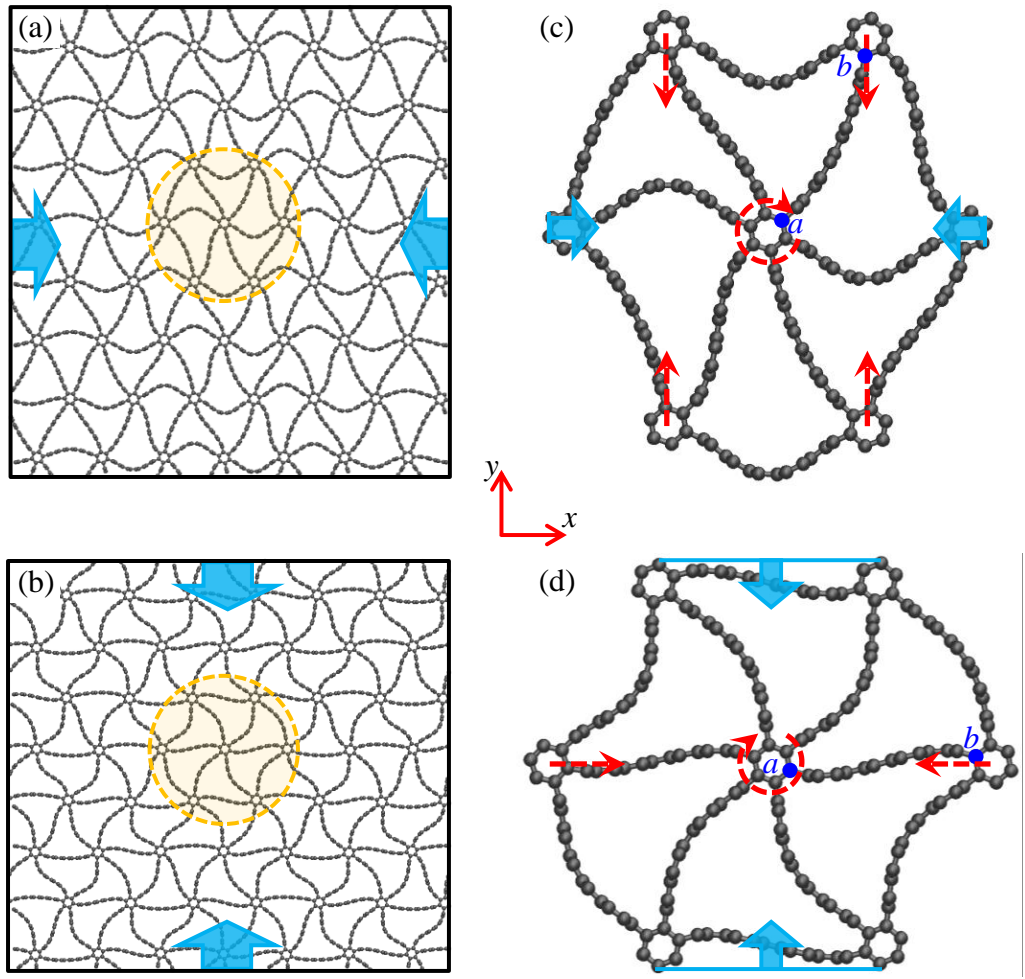

Figure 4. Structures of CFs after buckling. Left: typical structures of the buckled CFs when they are uniaxially compressed along (a) *x* and (b) *y* directions. Right: representative unit cells taken from the buckled CFs under uniaxial compressive loads in (c) *x* and (d) *y* directions. After buckling, the axial compressive loads (blue arrows) in CFs will induce the rotation of the central hexagon carbon ring and thus induce the contraction in transverse directions (red arrows), which finally results in the auxeticity in the buckled CFs.

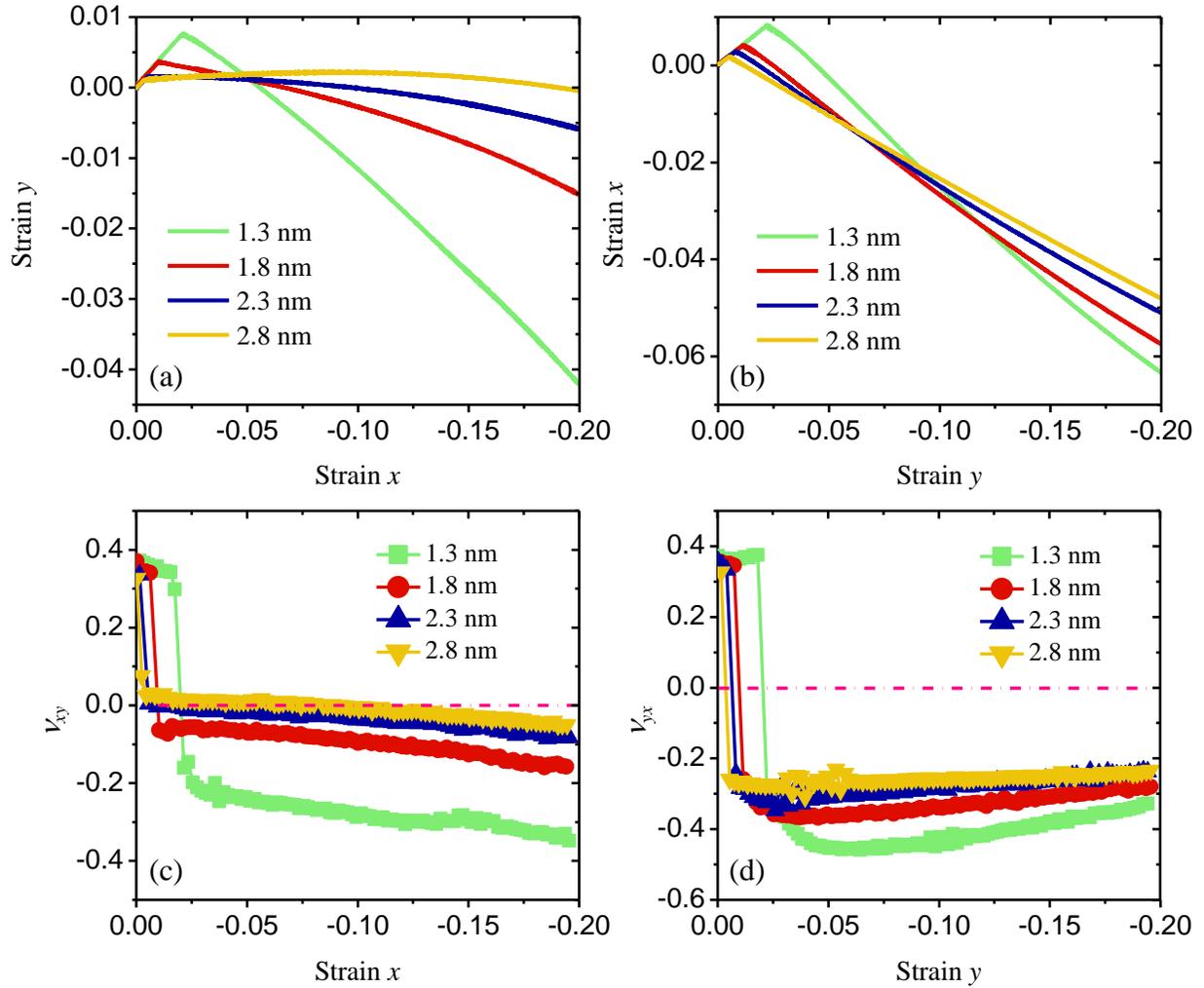

Figure 5. Poisson's ratios of CFs with different cell lengths. Top: the transverse strain versus the applied axial strain when CFs are uniaxially compressed along (a) *x* and (b) *y* directions. Bottom: Poisson's ratios of CFs compressed in (c) *x* and (d) *y* directions. The negative Poisson's ratio phenomenon initiates after CF buckles and retains over a wide range of applied compression.

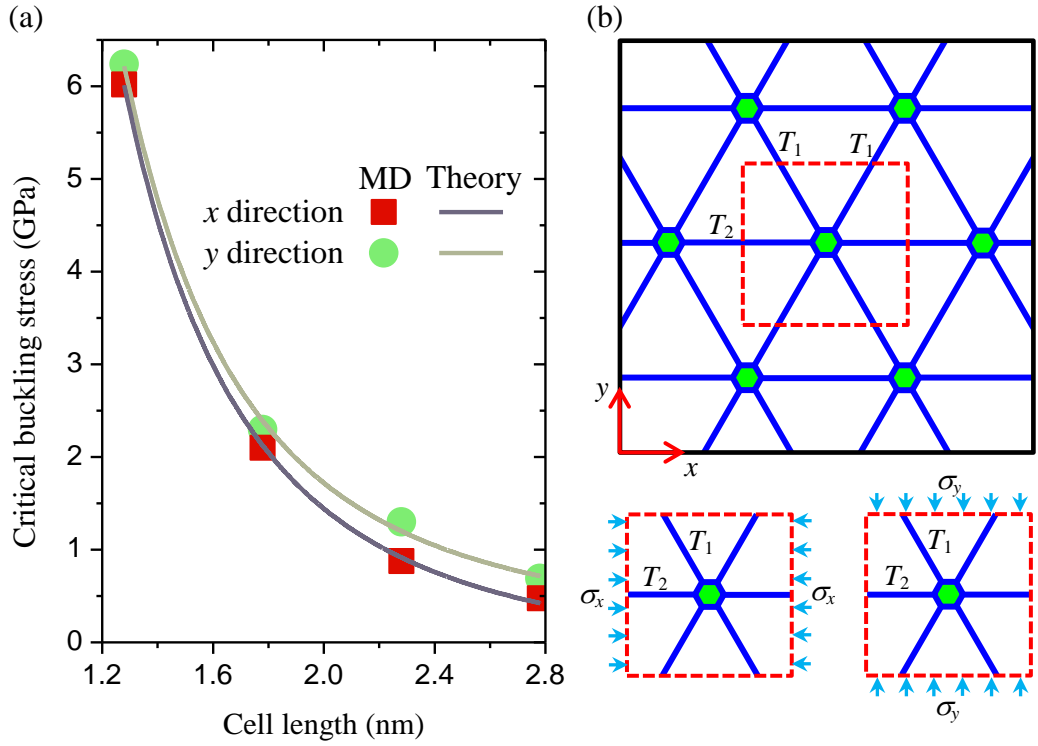

Figure 6. Critical buckling stress of CFs. (a) The critical buckling stress $\sigma_{cr}$ of CFs with different cell lengths $d$ when they are compressed in different directions. The critical buckling stress of CFs calculated from MD simulations can be well fitted by $\sigma_{cr} \propto d^{-3}$ predicted from the continuum model shown in (b). (b) The continuum model of CFs and the stress distribution in a representative unit cell of CFs when they are subjected to a uniaxial load in $x$ and $y$ directions, respectively.

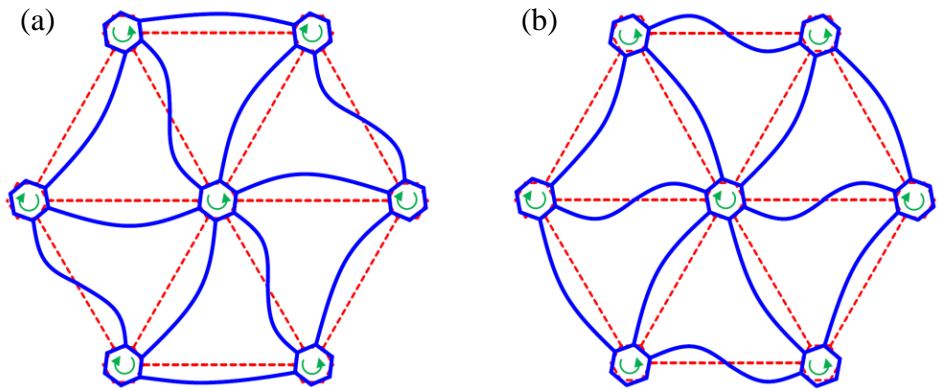

Figure 7. Buckling modes of a macroscopic structure having the geometry analogous to the present CFs. Here the solid blue lines present the deformed configurations after buckling, the dashed red lines denote the initial structures and the green arrows indicate the rotation of the hexagon ring after buckling. The buckling modes of the macroscopic structure under the uniaxial compression in (a) *x* and (b) *y* directions resemble the structures of the buckled CFs, denoting that similar negative Poisson's ratio behaviours may also exist in the macroscopic analogues of CFs.